\documentclass[preprint,preprintnumbers,amsmath,amssymb]{revtex4}
\usepackage{graphicx}
\usepackage{dcolumn}
\usepackage{bm}
\usepackage{epsfig}

\begin{document}


\title{Quantum mechanical analysis of the equilateral triangle billiard: 
periodic orbit theory and wave packet revivals}

\author{M. A. Doncheski} 
\affiliation{%
Department of Physics \\
The Pennsylvania State University\\
Mont Alto, PA 17237  USA \\
}

\author{R. W. Robinett} 
\affiliation{%
Department of Physics\\
The Pennsylvania State University\\
University Park, PA 16802 USA \\
}%


\begin{abstract}

Using the fact that the energy eigenstates of the equilateral triangle 
infinite well (or billiard) are available in closed form, we examine the 
connections between the energy eigenvalue spectrum and the classical closed 
paths in this geometry, using both periodic orbit theory and the short-term 
semi-classical behavior of wave packets. We also discuss wave packet revivals 
and show that there are exact revivals, for all wave packets, 
at times given by 
$T_{rev} = 9 \mu a^2/4\hbar \pi$ where $a$ and $\mu$ are the length of one 
side and the mass of the point particle respectively. We find additional cases 
of exact revivals with shorter revival times for zero-momentum wave packets 
initially located at special symmetry points inside the billiard.  Finally, we 
discuss simple variations on the equilateral 
($60^{\circ}-60^{\circ}-60^{\circ}$) triangle, such as the half equilateral 
($30^{\circ}-60^{\circ}-90^{\circ}$) triangle and other `foldings', which have 
related energy spectra and revival structures.

\end{abstract}

\maketitle

\section{\label{sec:level1} Introduction}

The study of the connections between the quantized energy eigenvalues of a 
bound state and the classical motions of the corresponding classical point 
particle has undergone something of a renaissance as the ability to 
experimentally probe the quantum-classical interface has dramatically 
improved. Theoretical methods such as periodic orbit theory 
\cite{gutzwiller_book}, \cite{gutzwiller_review}, \cite{semiclassical_book}, 
for example, provide very direct connections between the energy spectrum and 
the closed trajectories of the classical system. The time-dependence of fully 
quantum mechanical wave packet solutions of the Schr\"odinger equation also 
depends critically on the energy spectrum, both for the short-term 
semi-classical propagation as well as for longer-term, purely quantum 
mechanical effects such as wave packet revivals.  Such revival phenomena have 
been observed in a wide variety of physical systems, especially in Rydberg 
atoms \cite{revival_review}, and calculations exist for many other systems 
\cite{other_revivals}.

Two-dimensional billiard systems have provided easily visualizible examples 
relevant for both types of analyses. For example, the periodic orbit theory 
analysis of square/rectangular \cite{gutzwiller_paper} and spherical/circular 
\cite{balian_and_bloch} geometries 
were among the first performed, while on the 
experimental side, the energy level structure and statistics of microwave 
cavities \cite{microwave} (modeling 2D billiards of arbitrary shape) have been 
probed to test both periodic orbit theory and to obtain statistical evidence 
of chaotic behavior.  Measurements of conductance fluctuations in ballistic 
microstructures \cite{microstructures} have been tentatively used to identify 
frequency features in the power spectrum with specific closed orbits in a 
circular (and stadium) billiard.  More recently, the realization of 
atom-optics billiards \cite{atom_optics}, with ultra-cold atoms in arbitrary 
shaped 2D boundaries confined by optical dipole potentials, has allowed the 
study of various chaotic and integrable shapes such as the stadium, ellipse 
and circle, again for short-term, semi-classical propagation.  

The one- \cite{one_d_square_wells} and two-dimensional \cite{bluhm_2d}, 
\cite{other_2d} square wells, with their integral energy spectra, have 
provided simple examples of exact wave packet revivals, wherein initially 
localized states which have a short-term, quasi-classical time evolution, can 
then spread significantly over several orbits, only to reform later in the 
form of a quantum revival in which the spreading reverses itself, the wave 
packet relocalizes, and the semi-classical periodicity is once again evident.  
Circular billiards have also been recently studied \cite{robinett_pra} where 
approximate revivals have been found to be  present under some circumstances. 

One billiard system which has not been analyzed in as much detail is the case 
of the equilateral (or $60^{\circ}-60^{\circ}-60^{\circ}$) triangle quantum 
infinite well or billiard.  If one thinks of the square billiard as an $N=4$ 
regular polygon, with the circle being the $N \rightarrow \infty$ limit, the 
$N=3$ case of an equilateral triangular billiard is an obvious extension.  
Since the exact 
bound state wavefunctions for the corresponding infinite well problem have 
been available for some time \cite{canadian}, \cite{berry}, 
\cite{math_methods}, our purpose here will be to apply these known results and 
analyze the quantum mechanics of this system in the context of periodic orbit 
theory as well as wave packet revivals. Given how few quantum systems, 
especially billiard geometries, have closed-form solutions, it is very useful 
to have another such example analyzed in as much detail as possible. In 
addition, the energy eigenvalues for this case, described by simple quadratic 
powers of quantum numbers, just as for the 2D square billiard, will be seen to 
yield exact quantum revivals, providing another interesting `testbed' for 
possible future experimental tests. 

In Section~II, we review the solutions of the equilateral 
triangular well, wavefunctions 
and energy eigenvalues, and make some comments about simple variations 
resulting from various `foldings' of the standard equilateral triangle 
billiard. We then perform a periodic orbit theory analysis of this system in 
Sec.~III and discuss the exact wave packet revivals present in this system in 
Sec.~IV, as well as visualizing the short-term semiclassical propagation wave 
packets and their relationship to classical closed orbits.

\section{Energy eigenfunctions and eigenvalues}

\subsection{Results for the equilateral triangle billiard}

The energy eigenvalues and position-space wave functions for a particle of 
mass $\mu$ in an equilateral triangular infinite well (or billiard) of side 
$a$ have been derived, in a variety of different contexts, by at least three 
different groups \cite{canadian}, \cite{berry}, \cite{math_methods}.  For 
definiteness, we will assume a triangular billiard with vertices located at 
$(0,0)$, $(a/2,\sqrt{3}a/2)$, and $(-a/2,\sqrt{3}a/2)$, as shown in Fig.~1.  
The resulting energy spectrum is given by
\begin{equation}
E(m,n) = \frac{\hbar^2}{2\mu a^2} \left(\frac{4 \pi}{3}\right)^2
\left( m^2 + n^2 - mn\right)
\label{energy_eigenvalues}
\end{equation}
for integral values of $m,n$,  with the restriction that $m \geq 2n$.  
(In what 
follows, we will use the notations of Ref.~\cite{berry} for the energies and 
wavefunctions: we use $\mu$ for the particle mass to avoid confusion with 
various quantum numbers.)  For the case of $m > 2n$, there are two degenerate 
states with different symmetry properties \cite{berry} which can be written in 
the forms
\begin{eqnarray}
\psi_{(m,n)}^{(-)}(x,y) & = &
\sqrt{\frac{16}{a^2 3\sqrt{3}}}
\left[
\sin\left(\frac{2\pi (2m-n)x}{3a}\right) 
\sin\left(\frac{2\pi ny}{\sqrt{3}a}\right) 
\right. \nonumber  \\ 
& & 
\qquad \qquad
- 
\sin\left(\frac{2\pi (2n-m)x}{3a}\right) 
\sin\left(\frac{2\pi my}{\sqrt{3}a}\right) 
\label{odd_wavefunctions}
\\
& &
\qquad \qquad 
\left.
- 
\sin\left(\frac{2\pi (m+n) x}{3a}\right)
\sin\left(\frac{2\pi (m-n) y}{\sqrt{3}a}\right)
\right] \nonumber 
\end{eqnarray}
and
\begin{eqnarray}
\psi_{(m,n)}^{(+)}(x,y) & = &
\sqrt{\frac{16}{a^2 3\sqrt{3}}}
\left[
\cos\left(\frac{2\pi (2m-n)x}{3a}\right) 
\sin\left(\frac{2\pi ny}{\sqrt{3}a}\right) 
\right. \nonumber \\ 
& & 
\qquad \qquad 
- 
\cos\left(\frac{2\pi (2n-m)x}{3a}\right) 
\sin\left(\frac{2\pi my}{\sqrt{3}a}\right) 
\label{even_wavefunctions}  \\
& &
\qquad \qquad 
\left.
+
\cos\left(\frac{2\pi (m+n) x}{3a}\right)
\sin\left(\frac{2\pi (m-n) y}{\sqrt{3}a}\right)
\right] \, .
\nonumber 
\end{eqnarray}
Extending earlier results, we have here included the correct normalizations, 
since we will eventually expand Gaussian wave packets in such eigenstates.

For the special case of $m=2n$ there is a single non-degenerate state for each 
$n$, given by 
\begin{equation}
\psi_{(2n,n)}^{(o)}(x,y) = 
\sqrt{\frac{8}{a^2 3\sqrt{3}}}
\left[
2\cos\left(\frac{2\pi nx}{a}\right) \sin\left(\frac{2\pi n y}{\sqrt{3}a}\right)
- \sin\left(\frac{4 \pi ny}{\sqrt{3}a}\right)
\right] \, .
\label{special_wavefunctions}
\end{equation}
Clearly these states satisfy
\begin{equation}
\psi_{(m,n)}^{(\pm)}(-x,y) = 
\pm \psi_{(m,n)}^{(\pm)}(x,y)
\qquad
,
\qquad
\psi_{(m,n)}^{(o)}(-x,y) = 
+ \psi_{(m,n)}^{(o)}(x,y)
\end{equation}
and the $\psi_{(m=2n,n)}^{(\pm)}(x,y)$ states also satisfy 
\begin{eqnarray}
\psi_{(m=2n,n)}^{(+)}(x,y) & =& \sqrt{2} \psi_{(2n,n)}^{(o)}(x,y) \\
\psi_{(m=2n,n)}^{(-)}(x,y) & =& 0 \, .  \nonumber 
\end{eqnarray}
The authors of Ref.~\cite{berry} note that the 
$\psi^{(\pm)}_{(m,n)}(x,y)$ states also satisfy such relations as
\begin{equation}
\psi_{(m,m-n)}^{(\pm)}(x,y) = \pm \psi_{(m,n)}^{(\pm)}(x,y)
\qquad
\mbox{and}
\qquad
\psi_{(n,m)}^{(\pm)}(x,y) = - \psi_{(m,n)}^{(\pm)}(x,y)
\label{relationships}
\end{equation}
and several other similar ones. The extremely symmetric character of the 
$(2n,n)$ states can also be seen by the fact that it can be rewritten in the 
form 
\begin{equation}
\psi_{(2n,n)}^{(o)}(x,y)
= \frac{8\sqrt{2}}{3^{3/4}a}
\sin\left(\frac{2\pi n y}{\sqrt{3}a}\right)
\sin\left(\frac{\pi n (y-\sqrt{3}x)}{\sqrt{3}a}\right)
\sin\left(\frac{\pi n (y+\sqrt{3}x)}{\sqrt{3}a}\right)
\, . 
\end{equation}

These wavefunctions satisfy the Schr\"odinger equation with the energy 
eigenvalues in Eqn.~(\ref{energy_eigenvalues}),  as well as vanishing on the 
boundaries defined by the lines $y = \pm \sqrt{3}x$ and  $y = \sqrt{3}a/2$.  
The 
eigenfunctions corresponding to different eigenvalues can be shown to be 
orthogonal, while the two with the same eigenvalues 
($\phi_{(m,n)}^{(\pm)}(x,y)$) are obviously so due to 
their $x \rightarrow -x$ parity properties. The authors of both 
Ref.~\cite{canadian} and \cite{berry} argue that the solutions above 
also form a  complete set of  states by explicitly calculating the Weyl 
area rule for the energy level density
\begin{equation}
\rho_{0}(E) = \frac{A}{4\pi} \left(\frac{2\mu}{\hbar^2}\right)
- \frac{L}{8\pi} \sqrt{\frac{2\mu}{\hbar^2 E}}
\label{energy_level_density}
\end{equation}
and showing that this relation is saturated by the energy spectrum in 
Eqn.~(\ref{energy_eigenvalues}). While we will focus on the case of the 
equilateral triangle billiard, it is useful to note that several simple 
variations on this system (see below) can also be derived from these results.

We note that these solutions, while derived in the context of the explicit 
triangular area shown in Fig.~1 (shown in bold), are also solutions, when 
extended by reflections, in the other triangular regions (flipped in sign) and 
hence over the complete hexagonal area including reflections in the horizontal 
axis. (This connection, in fact, was used in the explicit construction of the 
solutions in one approach \cite{berry}.) Thus, these solutions (with their 
integrally quantized energies, will also form a subset of the solutions for 
the hexagonal ($N=6$ regular polygonal) billiard.

\subsection{Comparison to the square and circular billiards}

The degeneracy pattern of energy eigenvalues for the equilateral triangle 
billiard is very similar to that for the square ($N=4$ polygon) and circular 
($N=\infty$ polygon) systems. For the square billiard (of side $a$), the 
energy eigenvalues are given by
\begin{equation}
E(m,n) = \frac{\hbar^2 \pi^2}{2\mu a^2} \left( m^2 + n^2\right)
\end{equation}
with eigenfunctions given by
\begin{equation}
\psi_{(m,n)}(x,y) = u_{m}(x) u_{n}(y)
\end{equation}
with 
\begin{equation}
u_n(x) = \sqrt{\frac{2}{a}} \sin\left( \frac{n\pi x}{a} \right)
\end{equation}
and any integral $m,n \geq 1$ allowed. For $m=n$, there is a single state, 
while for $m \neq n$ there is a two-fold degeneracy. The analogs of the 
degenerate $(\pm)$ states of the equilateral triangle billiard for the square 
infinite well are given by
\begin{eqnarray}
\psi_{(m,n)}^{(-)}(x,y) & = & \frac{1}{\sqrt{2}} \left[u_n(x)u_m(y) - u_m(x)u_n(y)\right] 
\qquad
(m \neq n) \nonumber  \\ 
\psi_{(m,n)}^{(+)}(x,y) & = & \frac{1}{\sqrt{2}} \left[u_n(x)u_m(y) + u_m(x)u_n(y)\right] 
\qquad
(m \neq n) \\
\psi_{(n,n)}^{(o)}(x,y) & = & u_{n}(x)u_{n}(y) 
\, . \nonumber 
\end{eqnarray}
This form is useful since it allows one to discuss the energy eigenvalues and 
eigenfunctions of the $45^{\circ}-45^{\circ}-90^{\circ}$ triangle 
\cite{other_berry} billiard formed by `folding' the square along a diagonal 
\cite{robinett_jmp},  since the $\psi_{(m,n)}^{(-)}(x,y)$ satisfy the 
appropriate boundary condition along the new hypotenuse. Additional foldings 
along diagonals are also possible \cite{robinett_jmp} and the energy spectrum 
can be used to analyze these cases as well.

A similar pattern of degeneracies, also useful for analyzing `half-well' 
problems,  is found in the circular well \cite{robinett_jmp}, 
\cite{robinett_ajp}. In this case, the wavefunctions in polar coordinates can 
be written in the form
\begin{equation}
\psi(r,\theta) = \left[N_r J_{|m|}(kr)\right]\left[N_{\theta}
e^{im\theta}\right]
\qquad
\mbox{where $m = 0, \pm 1, \pm 2,$ ...}
\end{equation}
where the radial wavefunctions are normalized via 
\begin{equation}
\int_{0}^{R} \left[N_r J_{|m|}(kr)\right]^2 \,r\,dr = 1
\, . 
\end{equation}
The states corresponding to non-negative values of $+m$ and $-m$ are 
degenerate. A standard set of angular wavefunctions can be written as 
\begin{equation}
\int_{0}^{2\pi}\left|N_{\theta} e^{im\theta} \right|^2\,d\theta
= 1
\qquad
\mbox{or}
\qquad
N_{\theta} = 1/\sqrt{2\pi}
\, ,
\end{equation}
but for the purposes of visualizing the solutions of bound state problems, 
real wavefunctions are also possible, given by
\begin{equation}
    \Theta_{|m|}(\theta) = \left\{ \begin{array}{ll}
               \cos(m\theta)/\sqrt{\pi} &  \quad m \geq 1 \\ 
               \sin(m\theta)/\sqrt{\pi}  &  \quad m \geq 1 \\ 
               1/\sqrt{2\pi}  &  \quad m = 0
                                \end{array}
\right.
\, . 
\end{equation}
Thus, there is a doubly degenerate set of states labeled by $(m>1,n_r)$ for 
$m > 1$, and a single set of states (all of the $s$-state wavefunctions) with 
$(m=0,n_r)$. One of the $(m>1,n_r)$ sets can then be used to describe the 
energies of the `half-circular' billiard, as either the $\cos(m\theta)$ or 
$\sin(m \theta)$ solutions (or some linear combination thereof) can be used to 
satisfy the boundary condition along an infinite wall on any diameter, giving 
a `folded half-well'.

\subsection{Variants on the equilateral triangle}

Using the explicit wavefunctions for equilateral triangle, it is easy to 
construct solutions of the `half-triangle' (namely the 
$30^{\circ}-60^{\circ}-90^{\circ}$ triangle) obtained by folding the 
equilateral triangle along one bisector: the odd-parity solutions in 
Eqn.~(\ref{odd_wavefunctions}), for example, explicitly satisfy the new 
boundary condition along the bisector given by $x=0$.  The resulting energies 
are still given by Eqn.~(\ref{energy_eigenvalues}), but now with only a single 
copy of the $m>2n$ states allowed, which  agrees with earlier analyses 
\cite{canadian}, \cite{other_berry}. The wavefunctions in the new `half-well' 
are identical to the $\psi_{(m,n}^{(-)}(x,y)$ except with an additional 
normalization factor of $\sqrt{2}$ to account for the geometrical `footprint' 
of the smaller billiard.

The three lowest-lying such states, corresponding to $(m,n)$ quantum numbers 
and energy values of $(3,1), (4,1)$, and $(5,2)$ and $7E_0, 13E_0$, and 
$19E_0$ respectively (where $E_0 \equiv (\hbar^2/2\mu a^2)(4\pi/3)^2$) are 
shown on the top line of Fig.~2. 

It is easy to shown that if one lets $(m',n') = (2m,2n)$ then all of the 
energy eigenvalues are multiplied by the same factor of $4$ and that there are 
nodal lines which split up each `half-triangle' into four equal parts 
corresponding to a simple folding. The corresponding states for this subset 
of quantum numbers are also shown in Fig.~2 (in the bottom row).  This 
possibility is part of a general pattern where the mapping 
$(m',n') = (fm,fn)$ yields $f^2$ `copies' of the $(m,n)$ wavefunctions (for 
both the even and odd states) contained inside the original triangular 
billiard; one can see this more explicitly by noting that
\begin{equation}
\psi_{(fm,fn)}^{(\pm)}(x,y;a) = \psi_{(m,n)}^{(\pm)}(x,y;a/f)
\, . 
\end{equation}

Slightly less trivially, the quantum number mapping given by 
$(m',n') = (2m-n,m-2n)$ gives states which are a factor of $3$ times in 
energy, with nodal lines along the perpendicular bisectors, as shown in (the 
middle row of) Fig.~2. A second application of this map gives 
$(m'',n'') = (3m,3n)$ and a resulting common increase by a factor of $9$, as 
mentioned above. Both sets of results can be then be easily understood in 
terms of repeated foldings of the `half-triangle' onto itself, with 
increasingly small areas and larger energy eigenvalues scaling in the 
appropriate manner, just as in Ref.~\cite{robinett_jmp} for diagonal foldings 
of the square billiard. For this case, we might note that the odd 
wavefunctions explicitly satisfy the relation 
\begin{equation}
\psi_{(2m-n,m-2n)}^{(-)}(x,y;a) = -\psi_{(m,n)}^{(-)}(y,x;a/3)
\, .
\end{equation}

Other simple linear transformations of quantum numbers of the form 
$(m',n') = (\alpha m + \beta n, \gamma m + \delta n)$ also return allowed 
quantum numbers in the triangular well, corresponding to higher energy states, 
but without the same obvious geometrical significance.  A few examples of such 
mappings are discussed briefly in the Appendix.

\section{ Periodic orbit theory analysis}

\subsection{General background}

As mentioned above, the notion of density of energy levels has already been 
applied to the solutions of triangular billiard to the extent that the 
averaged energy level density, $\rho_{0}(E)$, in 
Eqn.~(\ref{energy_level_density}), has been evaluated to confirm completeness, 
as in Refs.~\cite{canadian} and \cite{berry}. It is perhaps then an obvious 
next step to expand on this type of approach by using a periodic orbit theory 
analysis \cite{gutzwiller_book}, \cite{gutzwiller_review}, 
\cite{semiclassical_book}. In this context, the energy level density is 
naturally split into the smooth, slowly varying part ($\rho_{0}(E)$) and a 
second, oscillatory term which is  dominated (in a saddle point approximation 
of the path integral) by classical orbits whose  actions,  $S_{\gamma}(E)$, 
correspond  to periodic orbits or closed paths.
Specifically, using the 
quantized energy eigenvalues, labeled collectively as $E_n$, one writes
\begin{equation}
\sum_{n=1}^{\infty}\delta(E-E_n) \equiv \rho(E) = \rho_{0}(E) +
\sum_{p=1}^{\infty}\sum_{\gamma} \rho_{\gamma,p} 
\cos\left[p\left(\frac{S_{\gamma}(E)}{\hbar} - \phi_{\gamma}\right)\right]
\end{equation}
where each periodic orbit is characterized by a label $\gamma$ and $p=1,2,...$ 
runs over all possible repetitions of such trajectories.  For billiard 
systems, this expression can be Fourier transformed and simplified in terms of 
\begin{equation}
\rho(L) \equiv   \sum_{n=1}^{\infty}
\int_{-\infty}^{+\infty} \delta(k-k_n)\, e^{ikL}\, dk
= \sum_{n=1}^{\infty} e^{ik_nL}
\label{ffourier_sum}
\end{equation}
to give 
\begin{equation}
\rho(L) =  \sum_{p=1}^{\infty} \sum_{\gamma}
\rho_{\gamma,p} \delta(L - pL_{\gamma})
\, . 
\end{equation}
Thus, an evaluation of $\rho(L)$, using the quantized $k_n$ values given by 
the $E_{n} = \hbar^2k_{n}^2/2\mu$, will exhibit a series of $\delta$-function 
like `spikes' at multiples of the lengths of the `primitive' closed paths, 
{\it i.e.}, at $L = pL_{\gamma}$.  For numerical studies, an evaluation of
\begin{equation}
\rho_{N}(L) = \sum_{n=1}^{N} e^{ik_nL}
\label{finite_sum}
\end{equation} 
for large, but finite $N$ will give a series of sharp peaks corresponding to 
the closed orbits, with the smooth $\rho_{0}(E)$ piece corresponding to a 
large, but uninteresting feature at $L=0$.

\subsection{Periodic orbit theory analysis of the equilateral triangle 
billiard}

For a numerical evaluation of $\rho_{N}(L)$, we require the quantized 
wavenumbers obtained from the energy spectrum in 
Eqn.~(\ref{energy_eigenvalues}), namely
\begin{equation}
k_{(m,n)}a = \left(\frac{4 \pi}{3}\right)\sqrt{m^2 + n^2 - mn}
\end{equation}
so that
\begin{equation}
\left\{k_{(m,n)}; m/2 \geq n \geq 1 \right\} =
\left\{7.255, 11.082, 11.082, 14.510, 15.102, 15.102,...\right\}
\, , 
\end{equation}
where including the correct degeneracy, as we will see, is important.  An 
analysis of $\rho_{N}(L)$,  evaluated  using Eqn.~(\ref{finite_sum}), then 
requires knowledge of the possible closed paths or periodic orbits in the 
triangular billiard. 

A particularly easy way to visualize the possible classes of closed orbits 
in this geometry is 
by making use of the fact that the 2D plane can be tiled with a triangular 
lattice, as illustrated in Fig.~3. Consider a specific equilateral triangle 
billiard, in this case the one in the lower left hand corner.  By repeated 
foldings, the entire 2D plane can be covered, and `equivalent' triangles are 
shown with appropriate points identified.  Any point in the original triangle 
can then be connected to the corresponding point in another `identified' 
partner (as illustrated by the arrows) and the resulting path inside a 
specific triangular billiard can be obtained (visualized) by repeatedly 
folding the two triangles until they overlap.  (We note that this simple 
construction has been used for the square billiard \cite{annular_billiard} to 
obtain the path lengths for the closed orbits.)  In Fig.~4 we illustrate 
several cases corresponding to different starting points within the triangle, 
yielding identical overall path lengths, so long as the initial angles are the 
same. We note that, just as for the square and circular billiards, there are 
an infinite number of possible closed orbits of the same path length, here 
corresponding to different initial locations of the orbit. For example, the 
trajectories shown in Figs.~4(a) and (b) both correspond to path lengths of 
$L = 3a$ (and multiples thereof), while those in Figs.~4(d), (e), and (f) all 
yield path lengths of $\sqrt{3}a$.  Fig.~4(c) shows a unique (isolated) orbit 
which, because it closes on itself in a very symmetric way, corresponds to a 
single possible closed orbit with path lengths given by multiples of $3a/2$.

The general expression for the allowed closed orbit lengths can then be 
obtained using this construction. For example, without loss of generality, we 
can find the path lengths between a single corner  in the initial triangle in 
Fig.~3 (say the corner at $(0,0)$, labeled by the circle) and any other 
corresponding corner. For those points corresponding to triangles which have 
been folded into `even' rows, we can, for example,  use the points
\begin{equation}
x_{i_1} = i_1 (3a)
\qquad
\mbox{and}
\qquad
y_{j_1} = j_1(\sqrt{3}a)
\end{equation}
which gives
\begin{eqnarray}
d(i_1,j_1) & \equiv  & \sqrt{(x_{i_1}-0)^2 + (y_{j_1}-0)^2} \nonumber \\
& = & a\sqrt{9i_1^2 + 3j_1^2} = \frac{a}{2} \sqrt{9 (2i_1)^2 + 3 (2j_1)^2}
\, . 
\end{eqnarray}
This corresponds to initial angles (with respect to the horizontal) of
\begin{equation}
\tan(\theta) = \frac{y_{j_1}}{x_{i_1}} = \frac{j_1}{i_1\sqrt{3}}
\end{equation}
For points in triangles which have been folded into `odd' rows, we similarly 
have
\begin{equation}
x_{i_2} = (2i_2+1) \frac{3a}{2}
\qquad
\mbox{and}
\qquad
y_{j_2} = (2j_2+1) \frac{\sqrt{3} a}{2}
\end{equation}
giving closed path distances
\begin{eqnarray}
d(i_2,j_2) & \equiv & \sqrt{(x_{i_2}-0)^2 + (y_{j_2}-0)^2} \nonumber \\
& = & 
\frac{a}{2}\sqrt{9(2i_2+1)^2 + 3(2j_2+1)^2}
\end{eqnarray}
and initial angles
\begin{equation}
\tan(\theta) = \frac{y_{j_1}}{x_{i_1}} = \frac{(2j_2+1)}{(2i_2+1)\sqrt{3}}
\, . 
\end{equation}

We can combine these two cases into a single expression, namely
\begin{equation}
d(\overline{i},\overline{j}) = \frac{a}{2}
\sqrt{9 (\overline{i})^2 + 3 (\overline{j})^2}
\label{path_length_features}
\end{equation}
where $(\overline{i},\overline{j})$ are either {\it both} even or {\it both} 
odd. The corresponding angles are given by
\begin{equation}
\tan(\theta) = \frac{\overline{j}}{\overline{i}\sqrt{3}}
\, . 
\label{closed_orbit_angles}
\end{equation}
The path length features for these classes of closed orbits (and their 
recurrences) which have $L = d(\overline{i},\overline{j}) < 20a$, along with 
the corresponding $(\overline{i},\overline{j})$ labels and angles, are shown 
in Table~I. This construction finds all of the allowed closed orbits, except 
for the special case described by Fig.~4(c).

We can rewrite Eqn.~(\ref{path_length_features}) in a more symmetric form by 
defining
\begin{equation}
(p+q) \equiv \overline{i}
\quad
,
\quad
(p-q) \equiv \overline{j}
\qquad \quad
\mbox{or}
\qquad \quad 
p \equiv \frac{\overline{i}+\overline{j}}{2}
\quad
,
\quad
q \equiv \frac{\overline{i}-\overline{j}}{2}
\, . 
\end{equation}
Since $\overline{i},\overline{j}$ are either both even or both odd, the $p,q$ 
defined in this way are integer valued and we find that the path lengths can 
be written as 
\begin{equation}
d(p,q) = a\sqrt{3}\sqrt{p^2 + pq + q^2}
\label{other_path_length_features}
\end{equation}
and this form will be useful when we discuss how the closed or periodic orbits 
arise in the context of wave packet propagation in the next Section.

We then evaluate $\rho_{N}(L)$ (using the $N=1000$ lowest lying eigenvalues) 
and plot $|\rho_N(L)|^2$ versus $0 \leq L/a \leq 20$ in Fig.~5.  The path 
length features collected in Table~I are shown as vertical dotted lines and 
the vertical scale has been chosen to emphasize these features.  We note also 
that there are two much smaller features, indicated by arrows, corresponding 
to the isolated, single closed orbit trajectories at the first two odd 
multiples of $L/a = 3/2$, namely $L/a = 1.5$ (denoted by 
$(\overline{i},\overline{j}) = (2,0)'$) and $L/a = 4.5$ (for $3(2,0)'$).  Even 
recurrences are, of course, hidden under the standard 
($(\overline{i},\overline{j}) = (2,0)$) $L/a = 3.0$ orbits, while higher odd 
recurrences are close to other features corresponding to different orbits: for 
example the $5(2,0)'$ feature at $L/a=7.5$ is near a more prominent $(5,1)$ 
feature at $L/a=7.55$,  while  the $L/a = 10.5$, $7(2,0)'$ feature is shadowed 
by the $L/a = 10.53$ $(7,1)$ orbit.

\subsection{Periodic orbit theory analysis of the `half-triangle'}

We can easily extend the analysis of the equilateral triangle billiard to the 
`half-triangle' or $30^{\circ}-60^{\circ}-90^{\circ}$ triangle by using the 
reduced set of $(m>2n)$ energy eigenvalues corresponding to the 
$\psi_{(m,n)}^{(-)}(x,y)$ states which vanish along the $y$-axis as discussed 
in Section II~C and the evaluation of $\rho_{N}(L)$  proceeds just as above.  
The only substantiative changes in the allowed classical closed orbits can be 
visualized as has been done for various foldings of the square and circular 
wells in Ref.~\cite{robinett_jmp}. For all allowed orbits (save one special 
one), the introduction of an infinite wall along a bisector simply serves to 
reflect all trajectories back onto themselves inside the half-triangle, giving 
identical overall path lengths for closed orbits, as exemplified in Figs.~6(a) 
and (b) for two of the $(\overline{i},\overline{j}) = (1,1)$ orbits. In this 
case only, there is a second isolated path (shown in Fig.~6(c)) with half the 
expected path length, namely $L/a = \sqrt{3}a/2$ (and multiples thereof) which 
we will call the $(1,1)'$ orbit.

We thus expect to see in the $|\rho_{N}(L)|^2$ spectrum the same path length 
features as for the full-triangle, but with much smaller new features at odd 
multiples of $L/a = \sqrt{3}/2$, or 
$L/a \approx 0.866, 2.598, 4.330, 6.062, 7.794, 9.526,...$ and so forth.  We 
plot, in Fig.~7, $\rho_{N}(L)$ for both the full-well (dashed) and half-well 
(solid) cases, adjusting the overall scales to emphasize the similar shape, 
and we note the agreement at all of the standard, non-isolated path length 
features in Table~I, as well as for the isolated $L/a = 3/2$ $(2,0)'$ 
orbit.  For the 
half-well data, however, we also see clear evidence for all of the new 
isolated $(1,1)'$ orbit features (shown by arrows.)

\section{Time-dependence of Gaussian wave packets}

\subsection{General time-dependence of wave packet solutions}

The connections between the energy eigenvalue spectrum of a quantum system and 
the classical dynamics of particles can also be explored through the 
time-dependence of wave packet solutions of the Schr\"odinger equation.  For a 
localized wave packet in a system characterized by a single quantum number, 
$n$, an arbitrary wave packet solution can be written in the form
\begin{equation}
\psi(x,t) = \sum_{n=1}^{\infty} a_n \psi_n(x)e^{-iE(n)t/\hbar}
\, . 
\end{equation}
One then typically expands the energy eigenvalues (assuming integral values) 
about the central value, $n_0$, used in the construction of the wave packet 
to write
\begin{equation}
E(n) \approx E(n_0) + E'(n_0)(n-n_0) + \frac{1}{2}E''(n_0)(n-n_0)^2
+ \frac{1}{6}E'''(n_0)(n-n_0)^3 + \cdots
\end{equation}
in terms of which the classical period, revival, and superrevival times 
\cite{one_d_square_wells} are given respectively by
\begin{equation}
T_{cl} = \frac{2\pi\hbar}{|E'(n_0)|}
\qquad \quad
T_{rev} = \frac{2\pi \hbar}{|E''(n_0)|/2}
\qquad \quad
T_{super} = \frac{2\pi \hbar}{|E'''(n_0)|/6}
\, . 
\end{equation}

For systems with two quantum numbers \cite{bluhm_2d}, \cite{other_2d}, and 
energies labeled by $E(n_1,n_2)$, the patterns of both classical periodicities 
and revival times can be much richer as we will see below.  We first examine 
how the closed orbits for the equilateral triangle in 
Eqn.~(\ref{path_length_features}) arise in this formalism.

\subsection{Short-term, semi-classical periodicity in the equilateral
triangle billiard}

For a system, such as the 2D billiard, for which the quantized energies depend 
on two quantum numbers, $(m,n)$, there are two corresponding classical periods 
\cite{bluhm_2d}, \cite{other_2d} given by
\begin{equation}
T_{cl}^{(m)} = \frac{2\pi \hbar}{\partial E(m,n)/\partial m}
\qquad
,
\qquad
T_{cl}^{(n)} = \frac{2\pi \hbar}{\partial E(m,n)/\partial n}
\, . 
\end{equation}
Classical periodic orbits will arise when the ratio of these two times is a 
rational number, or equivalently when
\begin{equation}
pT_{cl}^{(m)} = T_{cl}^{(po)} = q T_{cl}^{(n)}
\label{closed_orbit_condition}
\end{equation}
which defines the classical period, $T_{cl}^{(po)}$, for such closed orbits.  
For the equilateral triangle billiard, with energies given by
\begin{equation}
E(m,n) =  \frac{\hbar^2}{2 \mu a^2}
\left(\frac{4\pi}{3}\right)^2
\left( m^2 + n^2 - mn\right)
\, ,
\label{energy_spectrum}
\end{equation}
we find that
\begin{equation}
T_{cl}^{(m)} = \left[\frac{9\mu a^2}{4\hbar \pi} \right]\frac{1}{(2m-n)}
\qquad
,
\qquad
T_{cl}^{(n)} = \left[\frac{9\mu a^2}{4\hbar \pi} \right] \frac{1}{(2n-m)}
\, . 
\end{equation}
The condition leading to closed orbits in Eqn.~(\ref{closed_orbit_condition}) 
can then be written as
\begin{equation}
\frac{(2m-n)}{(2n-m)} = \frac{T_{cl}^{(n)}}{T_{cl}^{(m)}}
= \frac{p}{q}
\qquad
\quad
\mbox{or}
\quad
\qquad
n = m \left(\frac{2p+q}{2q+p} \right)
\, .
\end{equation}
If we substitute this condition into the energy spectrum in 
Eqn.~(\ref{energy_spectrum}), as well as equating the quantum energies with 
the classical kinetic energy, $\mu v_0^2/2$, in the billiard (where $v_0$ is 
the classical speed) we are led to the association
\begin{equation}
\frac{1}{2} \mu v_0^2 
\longleftrightarrow
E(m,n) = \left(\frac{16 \pi^2}{9}\right) \left(\frac{\hbar^2}{2\mu a^2}\right)
\left[ m^2 + m^2\left(\frac{2p+q}{2q+p}\right)^2 
- m^2 \left(\frac{2p+q}{2q+p}\right) \right]
\end{equation}
or
\begin{equation}
\left(\frac{2\mu v_0a}{4\pi \hbar}\right)^2
= m^2 \left[\frac{3(p^2+pq + q^2)}{(2q+p)^2} \right]
\, . 
\end{equation}
This implies that 
\begin{equation}
m = \left(\frac{2\mu v_0a}{4\pi \hbar} \right) \frac{(2q+p)}{\sqrt{3}\sqrt{p^2 + pq + q^2}}
\qquad
,
\qquad
n = \left(\frac{2\mu v_0a}{4\pi \hbar} \right) \frac{(2p+q)}{\sqrt{3}\sqrt{p^2 + pq + q^2}}
\, .
\end{equation}
The period for classical, closed/periodic orbits is then given by 
Eqn.~(\ref{closed_orbit_condition}) as 
\begin{equation}
T_{cl}^{(po)} = pT_{cl}^{(m)}
= \frac{a \sqrt{3}\sqrt{p^2 + pq + q^2}}{v_0}
=
\frac{L(p,q)}{v_0}
\label{closed_orbit_period}
\end{equation}
where 
\begin{equation}
L(p,q) = d(p,q) = a\sqrt{3}\sqrt{p^2 + pq + q^2}
\end{equation}
and we have used the alternative form of the path lengths for periodic orbits 
in Eqn.~(\ref{other_path_length_features}).  (A similar analysis for the 
circular billiard can be found in Ref.~\cite{robinett_pra}, making use of the 
WKB approximation for the non-integral energies found there, while the 
corresponding derivation for the square billiard is entirely straightforward.)

To visualize this result, we make use of a specific form for Gaussian wave 
packets, namely
\begin{equation}
\psi(x,y;t=0) = \psi_0(x;x_0,p_{0x},b)\psi_0(y;y_0,p_{0y},b)
\label{initial_gaussian}
\end{equation}
where
\begin{equation}
\psi_0(x;x_0,p_{0x},b) = \frac{1}{\sqrt{b\sqrt{\pi}}} e^{ip_{0x}(x-x_0)/\hbar}
e^{-(x-x_0)^2/2b^2}
\label{gaussian_form}
\end{equation}
with a similar expression for $\psi_0(y;y_0,p_{0y},b)$.

The initial expectation values for the $x$ variables are given by
\begin{equation}
\langle x\rangle_0 = x_0
\qquad
,
\qquad
\langle x^2 \rangle_0 = x_0^2 + \frac{b^2}{2}
\qquad
,
\qquad
\Delta x_0 = \frac{b}{\sqrt{2}}
\end{equation}
and
\begin{equation}
\langle p_x\rangle_0 = p_{0x}
\qquad
,
\qquad
\langle p^2_x \rangle_0 = p_{0x}^2 + \frac{\hbar^2}{2b^2}
\qquad
,
\qquad
\Delta p_0 = \frac{\hbar }{\sqrt{2}b}
\end{equation}
with similar results for $y$.  The expectation value of total energy is 
\begin{equation}
\langle \hat{E} \rangle = \frac{1}{2m} \langle \hat{p}_x^2 
+ \hat{p}_y^2 \rangle
= \frac{1}{2m} \left[(p_{0x})^2 + (p_{0y})^2 + \frac{\hbar^2}{b^2}\right]
\, . 
\label{gaussian_energy}
\end{equation}

Just as with the allowed wave functions, wave packets in the billiard must 
satisfy the appropriate boundary conditions at the edges of the well.  
However, the simple Gaussian form above can be used to describe an initial 
state so long as the initial location, $(x_0,y_0)$, is well away from the 
edges of the potential well, so that a negligibly (exponentially) small error 
is made. In that case, such a Gaussian form can be easily and reproducibly 
expanded in terms of eigenstates, namely
\begin{eqnarray}
\psi(x,y;t=0) & = & \sum_{m>2n}^{\infty} \sum_{n=1}^{\infty}
\left(
a_{(m,n)}^{(+)} \psi_{(m,n)}^{(+)}(x,y)
+
a_{(m,n)}^{(-)} \psi_{(m,n)}^{(-)}(x,y)
\right) \nonumber \\
& & \qquad \qquad \qquad +
\sum_{n=1}^{\infty} a_{(2n,n)}^{(o)} \psi_{(2n,n)}^{(o)}(x,y)
\, . 
\label{expansion_coefficients}
\end{eqnarray}
For a sharply peaked Gaussian, we can (very accurately) approximate the 
required expansion coefficients by extending the required integration of 
$(x,y)$ over the triangular region to one covering all space, making use of 
the exact form of the wavefunctions.  Given the simple results in 
Eqns.~(\ref{odd_wavefunctions}), 
(\ref{even_wavefunctions},)
and (\ref{special_wavefunctions}), we can perform the necessary Gaussian 
integrals in closed form using 
\begin{eqnarray}
\int_{-\infty}^{+\infty}\, e^{ip_0(x-x_0)/\hbar}\, 
e^{-(x-x_0)^2/2b^2}\,\cos\left(\frac{Cx}{a}\right)\,dx
& = &
\frac{b\sqrt{2\pi}}{2}
\left[
e^{iCx_0/a} e^{-b^2(C/a + p_0/\hbar)^2/2} \right.
 \\
& & 
\qquad
\quad
\left. 
+
e^{-iCx_0/a} e^{-b^2(-C/a + p_0/\hbar)^2/2}
\right]
\nonumber
 \end{eqnarray}
and
\begin{eqnarray}
\int_{-\infty}^{+\infty}\, e^{ip_0(x-x_0)/\hbar}\, 
e^{-(x-x_0)^2/2b^2}\,\sin\left(\frac{Cx}{a}\right)\,dx
& = &
\frac{b\sqrt{2\pi}}{2i}
\left[
e^{iCx_0/a} e^{-b^2(C/a + p_0/\hbar)^2/2} \right.
 \\
& & 
\qquad
\quad
\left. 
-
e^{-iCx_0/a} e^{-b^2(-C/a + p_0/\hbar)^2/2}
\right] \, . 
\nonumber
 \end{eqnarray}
We can then construct wave packets with arbitrary initial locations 
$(x_0,y_0)$ inside the well (again, so long as they are well away from any 
edge, typically more than a few times $\Delta x_0 = \Delta y_0 = b/\sqrt{2}$) 
with arbitrary initial momenta, $(p_{0x},p_{0y})$, which we parameterize via 
$(p_{0}\cos(\theta),p_{0}\sin(\theta))$ to make connections with the specific 
angles characterizing closed orbits in Eqn.~(\ref{closed_orbit_angles}) and 
Table~I.

All quantum wave packets have a characteristic spreading time, so in order to 
exhibit behavior typical of a semi-classical regime, we want to have classical 
periodicities shorter than a typical time over which a Gaussian wave packet 
will spread significantly. For the free-particle form in 
Eqn.~(\ref{gaussian_form}), one can write
\begin{equation}
\Delta x_t = \Delta x_0 \sqrt{1 + (t/t_0)^2}
\qquad
\mbox{where}
\qquad
t_0 \equiv \frac{m b^2}{\hbar} = \frac{(2m)}{\hbar^2} (\Delta x_0)^2
\, . 
\end{equation}
The classical periods for closed orbits will be given by 
Eqn.~(\ref{closed_orbit_period}) and the two shortest such periods will be
\begin{equation}
T_{cl}^{(po)}(min) = \frac{ (\sqrt{3},3)a}{v_0} = 
\frac{(2\mu) a (\sqrt{3},3)}{2p_0}
\end{equation}
and we want to have $T_{cl}^{(po)}(min) < t_0$.  For purposes of illustration 
then, we have taken the following numerical parameters:
\begin{equation}
\hbar, 2\mu, a = 1
\quad
,
\quad
b = 1/10\sqrt{2}
\quad
\mbox{so that}
\quad
\Delta x_0 = \Delta y_0 = 0.05
\, , 
\quad
\mbox{and} 
\quad
p_0 = 1500.
\label{wave_packet_parameters}
\end{equation}
With these values, $t_0 = (0.05)^2 = 2.5 \times 10^{-3}$, while 
$T_{cl}^{(po)}(min) = (0.58, 1.00) \times 10^{-3}$ and one can see many 
classical periods for most of the lowest path length classical closed orbits 
before significant spreading occurs.

In order to probe whether a wave packet has returned to something close to its 
initial state after some expected classical period, it is standard to examine 
the autocorrelation function \cite{autocorrelation}, defined, for example, in 
one-dimension as
\begin{equation}
A(t) \equiv \int_{-\infty}^{+\infty}\, \psi^*(x;t)\, \psi(x;0)
= \sum_{n=1}^{\infty} |a_n|^2e^{iE_nt/\hbar}
\end{equation}
which also explicitly depends on the energy spectrum.  Evaluating $A(t)$ for 
our two-dimensional wave packets, we then plot, in Fig.~8, $|A(t)|$ versus 
$t/\tau$ (where $\tau \equiv a/v_0$) for Gaussian wave packets which are 
initially located at the geometric center of the triangular well 
($(x_0,y_0) = (0,\sqrt{3}a/3)$), and with the initial momentum in the 
$\theta$ direction: we then show results for various $\theta$ values in the 
range $0^{\circ} \leq \theta \leq 30^{\circ}$. (Because of the symmetry of the 
billiard, such plots will be symmetric about $\theta = 30^{\circ}$.)

We expect to see features in $A(t)$ at values of $\theta$ and classical 
periods $L(p,q)/v_0$ given in Table~I as we `scan' up in angle. The stars in 
Fig.~8 denote the location (in $(\theta,T_{cl}^{(po)})$ parameter space) where 
we would expect classical particles to return to their starting point in phase 
space and we do indeed see well-localized structures in $A(t)$ near each 
expected location. (The dotted curves which connect the `stars' representing 
the exact closed orbits are simply there to guide the eye.)  We see this same 
pattern if we initially localize the Gaussian wave packet to places in the 
well other than at the geometrical center, corresponding to the infinite sets 
of non-isolated classical orbits which all have the same path length.

One exception to this occurs for parameters near the initial location 
$(x_0,y_0) = (0,\sqrt{3}a/4)$ and with $\theta=0$ and the results for $A(t)$ 
for such this case are shown (on the bottom line) as the dotted peaks.  They 
coincide with the standard case, but also exhibit structure at classical path 
lengths given by odd multiples of $3a/2$ and are due to the special isolated 
$(\overline{i},\overline{j}) = (2.0)'$ path shown in Fig.~4(c).

Such wave packet solutions can also be constructed in the `half-well' case, 
using the appropriately normalized versions of the $\psi_{(m,n)}^{(-)}(x,y)$.  
As mentioned above, an initial Gaussian wave packet will be expandable (to 
exponential accuracy) in this way, provided it is contained within an 
appropriate `fiducial' area of the triangle, well away from any of the 
boundaries. We note that because such a wave packet would also be expandable 
in the original `full-well' (with coefficients given by 
Eqn.~(\ref{expansion_coefficients})), the expansion coefficients in the 
`half-well', $\tilde{a}_{(m,n)}^{(-)}$, will necessarily satisfy the relation
\begin{equation}
\tilde{a}_{(m,n)}^{(-)} =
\sqrt{2}a_{(m,n)}^{(-)} 
\end{equation}
Since the $\tilde{a}_{(m,n)}^{(-)}$ must saturate the complete probability for 
the `half-well' version of the wave packet, we must then have
\begin{equation}
1 = \sum_{m>2n}^{\infty} \sum_{n=1}^{\infty} |\tilde{a}_{(m,n)}^{(-)}|^2
= 2 
\sum_{m>2n}^{\infty} \sum_{n=1}^{\infty} |a_{(m,n)}^{(-)}|^2
\end{equation}
and the odd expansion coefficients (for such wave packets which are 
well inside one half of the original 
`full well') give exactly half of the total 
probability.  For wave packets in the `half-well', the only new closed orbit 
path length feature,  beyond those seen in Fig.~8,  arises from wave packets 
which are constructed to mimic the special isolated $(1,1)'$ 
orbit in Fig.~6(c).

\subsection{Long-term revival behavior}

The long-term behavior of wave packets can include the possibility of 
revivals, wherein initially localized states, after undergoing a number of 
semi-classical periods eventually spread out during a so-called collapsed 
phase, only to reform later into something very much like the initial state.  
The time scales governing these revivals, in a two-quantum number system are 
given by  \cite{bluhm_2d}, \cite{other_2d}
\begin{equation}
T_{rev}^{(m)} = \frac{2\pi \hbar}{(\partial^2 E/\partial m^2)/2}
\quad
,
\quad
T_{rev}^{(n)} = \frac{2\pi \hbar}{(\partial^2 E/\partial n^2)/2}
\quad
,
\quad
T_{rev}^{(m,n)} = \frac{2\pi \hbar}{(\partial^2 E/\partial m \partial n)}
\, . 
\end{equation}
For the equilateral triangle under consideration here, we have the simple 
result
\begin{equation}
T_{rev}^{(m)} 
= 
T_{rev}^{(n)} 
=
T_{rev}^{(m,n)} 
\equiv
T_{rev}
= \frac{9 \mu a^2}{4\hbar \pi}
\end{equation}
and exact revivals (just as for the 2D square well) are present in this 
system, with a single revival time guaranteed for any and all possible initial 
wave packets.

To visualize this phenomenon, we again make use of Gaussian wave packets, with 
the same parameters used above, but with vanishing momentum ($p_0 = 0$) where 
there would be no obvious classical periodicity.  We locate several such wave 
packets at initial positions (keeping away from the edges) along the $y$-axis, 
as shown in Fig.~9, and plot the autocorrelation function over a time equal to 
one revival time, $T_{rev}$.  In each case we do find an exact revival at the 
expected time, but we also notice special situations with exact revivals at 
shorter times, due to additional symmetries of the eigenfunctions, which 
result in the vanishing of certain expansion coefficients.  For example, for 
$(x_0,y_0) = (0,\sqrt{3}a/3)$ (geometric center of the well), one finds 
revivals at multiples of $T_{rev}/9$ since only expansion coefficients for 
which  $(m+n)$ is a multiple of $3$ are non-vanishing, due to the nodal 
pattern of the eigenstates superimposed on the circular symmetry of the 
$(p_{0x},p_{0y}) = (0,0)$ Gaussian wavefunction; in a similar fashion, if 
$(x_0,y_0) = (0,\sqrt{3}/4)$ (initial position half way down the bisector), 
one finds exact revivals at multiples of $T_{rev}/4$.  Even at these special 
points, if the momentum is made non-vanishing, only the true revivals at 
$T_{rev}$ remain intact, since the purely radially symmetric nature of the 
wavefunctions is lost. (Similar cases such as these are also seen in the 1D 
and 2D square well \cite{robinett_pra} for zero-momentum wave packets when 
obvious symmetries dictate that certain subsets of expansion coefficients 
vanish.)

Finally, since the `half-well' version of the triangular well also has 
integrally related energies, exact revivals are also present in the 
$30^{\circ}-60^{\circ}-90^{\circ}$ triangle, but without the special cases 
mentioned above present.

\section{Concluding remarks}

We have used the triangular billiard, with its simple energy spectrum, as a 
model system to probe periodic orbit theory and wave packet revivals.  
Similarly to the cases of the square ($N=4$ polygon) and circular 
($N \rightarrow \infty$ polygon) billiards, we find that 
for the equilateral triangular ($N=3$) infinite 
well the periodic orbit theory analysis agrees well with expectations, 
including special isolated orbits for various `folded' versions; for the case 
here, we find a special isolated orbit in the `full-well' version as well.  
The short-term, semi-classical propagation of quantum wave packets shows 
evidence, in the autocorrelation function, of the same closed orbits, 
including the isolated ones.

The equilateral triangle billiard also exhibits exact wave packet revivals 
(due to the integral nature of the energy eigenvalues), just like the 2D 
square billiard, as does the `half-well' version, for all wave packets.

The case considered here might also be used as a basis for a more systematic 
examination of the quantum energy spectrum of the $N=6$ polygon, namely the 
hexagonal billiard. The solutions here will form a subset of the allowed 
spectrum of the $N=6$ case and an examination of both the periodic orbits and 
wave packet revivals in such a system will require the complete enumeration of 
the energy eigenvalues, either in closed form, or perhaps using various 
approximation methods \cite{approximation}: having a subset of energy 
eigenvalues and wavefunctions for which the results are known exactly would, 
of course, provide a very useful `benchmark' for any numerical calculation.  
That case would also be another interesting example of a $\pi/N$ billiard 
\cite{hobson}.  Such a result would be useful in probing the pattern of wave 
packet revivals as one goes from the $N=3,4$ cases (where exact revivals are 
present) to the $N \rightarrow \infty$ (circular) case, where only approximate 
revivals exist.

\begin{acknowledgments}
The work of R. W. R. was supported, in part, by the National Science 
Foundation under Grant DUE-9950702. The work of M. A. D. was supported, in 
part, by the Commonwealth College of the Pennsylvania State University under a 
Research Development Grant (RDG).
\end{acknowledgments}

\appendix

\section{}

In Sec.~II(c) we exhibited several simple transformations of the $(m,n)$ 
quantum numbers of the equilateral triangular well which, when applied to 
allowed $(m,n)$ 
pairs, returned quantum numbers and corresponding energy eigenvalues 
appropriate  to higher energy states which, in turn, could be easily 
visualized in terms of simple `foldings' of the half-well. For example, the 
maps $(m',n') = (fm,fn)$ gave energies which are $f^2$ times larger, 
corresponding to $f^2$ smaller `copies' of the initial $(m,n)$ state inside 
the full triangular well. The less obvious mapping given by 
$(m',n') = (2m-n,m-2n)$ gave energies which were three times larger, 
corresponding to obvious `foldings' in thirds, as evident from the resulting 
wave function plots, as in Fig.~2.  

If we define a dimensionless function, $\epsilon(m,n)$, via
\begin{equation}
E(m,n) \equiv E_0\epsilon(m,n) =   E_0 (m^2 - mn + n^2)
\end{equation}
where $E_0 \equiv (\hbar^2/2\mu a^2)(4\pi/3)^2$, we can see that
\begin{equation}
\epsilon(fm,fn) = f^2 \epsilon(m,n)
\qquad
\mbox{and}
\qquad
\epsilon(2m-n,m-2n) = 3\epsilon(m,n)
\, .
\end{equation}
This second mapping is a special case of a general set of such mappings which 
return subsets of the allowed energy eigenvalue space.  If we pick any pair of 
integers, $(p,q)$, of the same form ($p>2q$) as the allowed quantum numbers 
themselves, and define the transformation
\begin{equation}
(m',n') = T_{(p,q)}[m,n]\equiv (pm-qn,(p-q)m - pn)
\end{equation}
one can easily confirm that
\begin{equation}
\epsilon(m',n') = [p^2 - pq + q^2](m^2 - mn + n^2)
= \epsilon(p,q) \epsilon(m,n)
\end{equation}
and we find all such energies in the spectrum of the equilateral triangle.
Two consecutive applications of the same transformation then give
\begin{equation}
(m'',n'') =   T_{(p,q)}\left[T_{(p,q)}[m,n]\right]
= ((p^2 - pq + q^2)m, (p^2 - pq + q^2)n)
\end{equation}
so that
\begin{equation}
\epsilon(m'',n'')  =  [\epsilon(p,q)]^2 \epsilon(m,n)
\, . 
\end{equation}
One can similarly show that two successive transformations with different 
labels, in either order,  give rise to equivalent sets of quantum numbers in 
the sense that the corresponding energies are identical, namely if
\begin{equation}
(m'',n'') = T_{(p,q)}[T_{(r,s)}[m,n]]
\qquad
\mbox{or}
\qquad
(m'',n'') = T_{(r,s)}[T_{(p,q)}[m,n]]
\end{equation}
both give
\begin{equation}
\epsilon(m'',n'') = [\epsilon(p,q)\epsilon(r,s)]\epsilon(m,n)
\, 
\end{equation}
which also appear in the original energy spectrum. 
We note that in order to see that the $(m'',n'')$ given in the two orderings 
are, in fact, identical, one sometimes must use the 
$(m,n) \leftrightarrow (m,m-n)$ equivalence in Eqn.~(\ref{relationships}).  
These transformations do not seem to have the same obvious geometrical 
significance as the $T_{(2,1)}(m,n)$ three way `folding' or the 
$(m',n') = (fm,fn)$ `copying'.

\newpage

\begin{flushleft}
{\large {\bf Tables}}
\end{flushleft}
\vskip 0.5cm

\begin{center}
\begin{tabular}{||c|r|l||c|r|l||} \hline
$(\overline{i},\overline{j})$   & $\theta$ (deg) & path lengths ($L/a < 20$) 
&
$(\overline{i},\overline{j})$   & $\theta$ (deg) & path lengths ($L/a < 20$) \\ \hline
$(2,0)$   & $0.0$  & $3.00, 6.00 ,9.00, 12.00, 15.00, 18.00$ &  $(11,5)$  & $14.7$ & $17.06$ \\ \hline
$(13,1)$  & $2.5$  & $19.52$ &  $(4,2)$   & $16.1$ & $6.24, 12.49, 18.73$ \\ \hline
$(11,1)$  & $3.0$  & $16.52$ &  $(9,5)$   & $17.8$ & $14.18$ \\ \hline
$(9,1)$   & $3.7$  & $13.53$ &  $(5,3)$   & $19.1$ & $7.94, 15.87$ \\ \hline
$(7,1)$   & $4.7$  & $10.53$ &  $(11,7)$  & $20.2$ & $17.58$ \\ \hline
$(12,2)$  & $5.5$  & $18.08$ &  $(6,4)$   & $21.1$ & $9.64, 19.28$ \\ \hline
$(5,1)$   & $6.6$  & $7.55, 15.10$ &  $(7,5)$   & $22.4$ & $11.36$ \\ \hline
$(13,3)$  & $7.6$  & $19.67$ &  $(8,6)$   & $23.4$ & $13.08$ \\ \hline
$(8,2)$   & $8.2$  & $12.12$ &  $(9,7)$   & $24.2$ & $14.80$ \\ \hline
$(11,3)$  & $8.9$  & $16.70$ & $(10,8)$  & $24.8$ & $16.5$ \\ \hline
$(3,1)$   & $10.9$ & $4.58, 9.16, 13.75, 18.33$ &  $(11,9)$  & $25.3$ & $18.24$ \\ \hline
$(13,5)$  & $12.5$ & $19.97$ & $(12,10)$ & $25.7$ & $19.97$ \\ \hline
$(10,4)$  & $13.0$ & $15.39$ & $(1,1)$   & $30.0$ & $1.73, 3.46, 5.20, 6.93, 8.66, 10.40,$  \\ \cline{1-3}
$(7,3)$   & $14.0$ & $10.82$ &           &        & $12.12, 13.86, 15.59, 17.32, 19.05$ \\ \hline
\end{tabular}
\end{center}

\begin{center}
Table~I. Path lengths (from Eqn.~(\ref{path_length_features}))
and initial angles (from Eqn.~(\ref{closed_orbit_angles}))
for closed orbits (and recurrences) with $L/a < 20$.
\end{center}

\newpage

\begin{flushleft}
{\large {\bf Figure Captions}}
\end{flushleft}
\vskip 0.5cm

\begin{itemize}
\item[Fig.\thinspace 1.] Geometry of the equilateral triangle billiard (shown 
in bold) with vertices at $(0,0)$, $(+a/2,\sqrt{3}a/2)$, and 
$(-a/2,\sqrt{3}a/2)$. 
\item[Fig.\thinspace 2.] Nodal patterns for the lowest-lying energy 
eigenstates for the `half triangular' well (using the 
$\psi_{(m,n)}^{(-)}(x,y)$ in Eqn.~(\ref{odd_wavefunctions})) on the top row. 
(The white (black) areas correspond to positive (negative) values of the 
wavefunction.) The bottom row corresponds to $(m',n') = (2m,2n)$ states which 
are a factor of $4$ larger in energy, while the middle row shows those related 
via $(m',n') = (2m-n,m-2n)$ which are a factor of $3$ higher in energy, 
illustrating various `foldings' of the half-well.
\item[Fig.\thinspace 3.] Geometrical visualization  leading to the 
construction of the path lengths corresponding to various closed orbits in the 
triangular billiard.
\item[Fig.\thinspace 4.] Closed orbits corresponding to several 
$(\overline{i},\overline{j}) = (1,1)$ cases (bottom row) with path lengths 
given by multiples of $\sqrt{3}a$ and for 
$(\overline{i},\overline{j}) = (2,0)$ cases (top row) with path lengths given 
by multiples of $3a$, except for (c) which is an isolated orbit (described as 
$(2,0)'$ in the text) with path length given by $3a/2$ (and multiples thereof.)

\item[Fig.\thinspace 5.] Plot of $|\rho_{N}(L)|^2$ versus $L/a$ for the 
triangular billiard, evaluated using the $1000$ lowest-lying energy 
eigenvalues. All of the expected path length features from 
Eqn.~(\ref{path_length_features}) and Table~I are indicated by vertical dotted 
lines. The small bumps indicated by arrows correspond to the isolated 
$(2,0)'$ orbits in Fig.~4(c) with path lengths given by multiples of $3a/2$.
\item[Fig.\thinspace 6.] Closed orbits in the `half-triangle' well for the 
$(\overline{i},\overline{j}) = (1,1)$ case. For (a) and (b), the path lengths 
are still given by $L(1,1) = \sqrt{3}a$ (and multiples thereof), but the 
`folding' induces a single isolated $(1,1)'$ orbit with $L = \sqrt{3}a/2$ as 
shown in (c).
\item[Fig.\thinspace 7.] Plot of $|\rho_{N}(L)|^2$ versus $L/a$, for the 
triangular billiard (dashed curve) and the `half-triangle' billiard (solid 
curve). Vertical dashed lines indicate the expected path length features in 
Table~I, while the vertical arrows indicate the small new features (at 
$L/a = \sqrt{3}/2$ and multiples thereof) present in the `half-triangle' due 
to the new, isolated closed orbit in Fig.~6(c).
\item[Fig.\thinspace 8.] Autocorrelation function, $|A(t)|$, versus $t/\tau$, 
(where $\tau \equiv a/v_0$) for Gaussian wave packets with the parameters 
discussed in Eqn.~(\ref{wave_packet_parameters}).  Wavepackets initially 
localized at $(x_0,y_0)$ corresponding to the geometric center of the triangle 
are given momenta  $(p_{0}\cos(\theta),p_{0}\sin(\theta))$ and $|A(t)|$ plots 
are shown for $\theta$ values in the range $(0^{\circ},30^{\circ})$.  The 
`stars' indicate the location of classical closed orbits (corresponding to the 
values in Table~I). The dotted $|A(t)|$ data for the $\theta = 0^{\circ}$ case 
corresponds to the initial wave packet localized at 
$(x_0,y_0) = (0,\sqrt{3}a/4)$ which corresponds classically to the isolated 
$(2,0)'$ orbit in Fig.~4(c).
\item[Fig.\thinspace 9.] Autocorrelation function, $|A(t)|^2$ versus $t$, for 
zero-momentum Gaussian wave packets initialized at various points along $x=0$ 
in the triangular well. There are exact revivals for all cases shown (where 
$|A(T_{rev})|=1$), but also special, shorter-period revivals at multiples of 
$T_{rev}/9$ and $T_{rev}/4$ for wave packets initially at positions of special 
symmetry in the well, namely at the geometric center $(0,\sqrt{3}a/3)$ (square 
box) and at $(0,\sqrt{3}a/4)$ (half way down the bisector.)
\end{itemize}

\begin{figure}
\epsfig{file=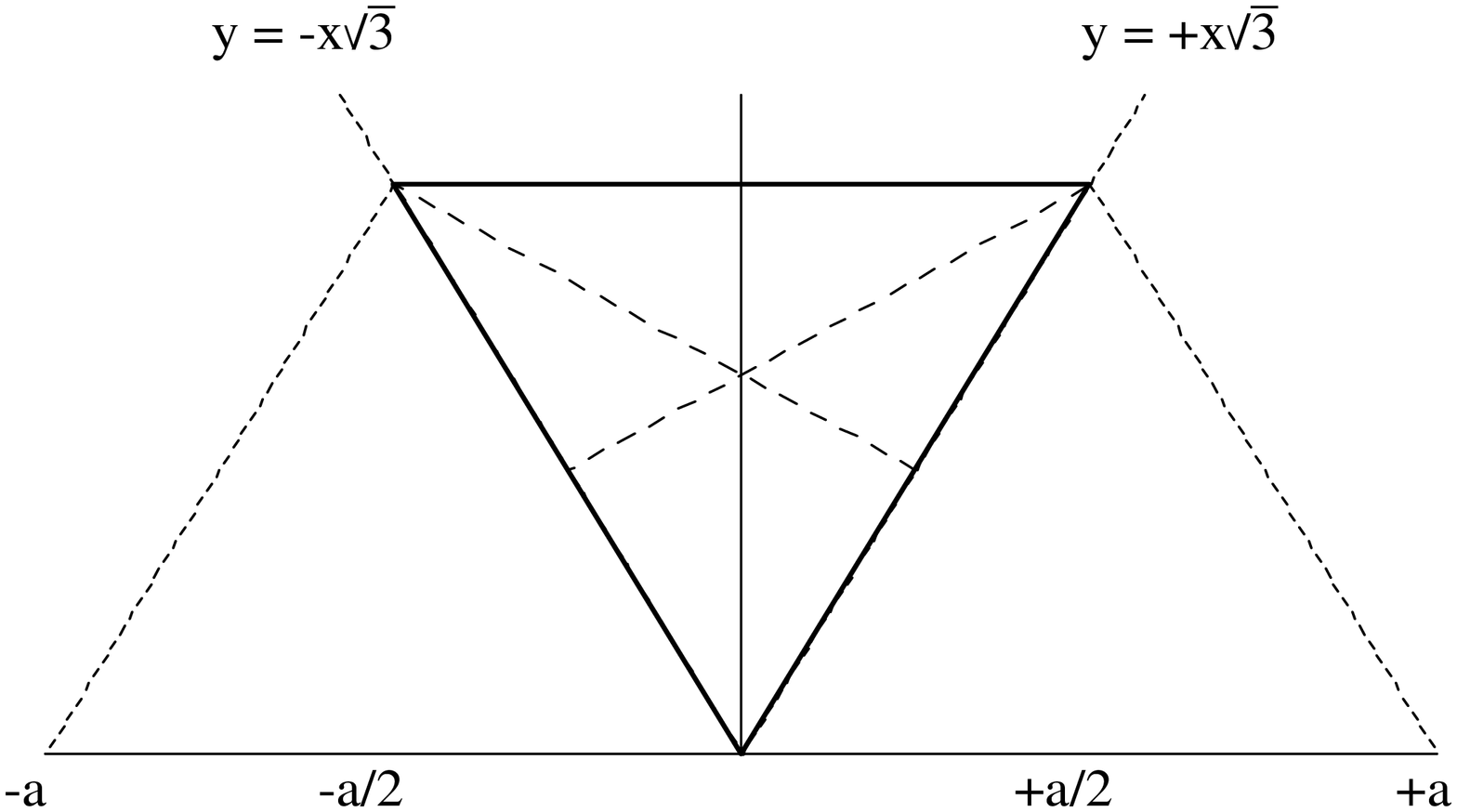,width=8cm,angle=270}
\caption{ }
\end{figure}

\newpage

\begin{figure}
\epsfig{file=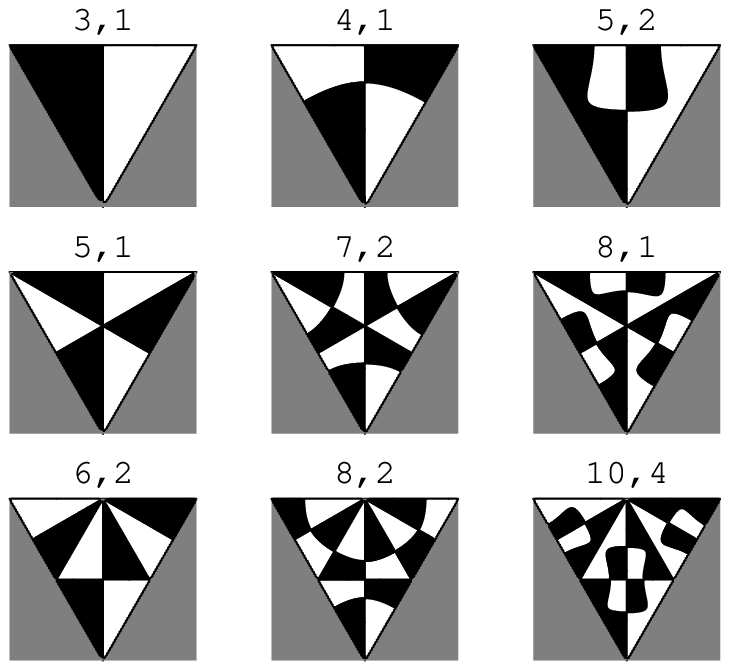,width=14cm,angle=0}
\caption{ }
\end{figure}

\newpage

\begin{figure}
\epsfig{file=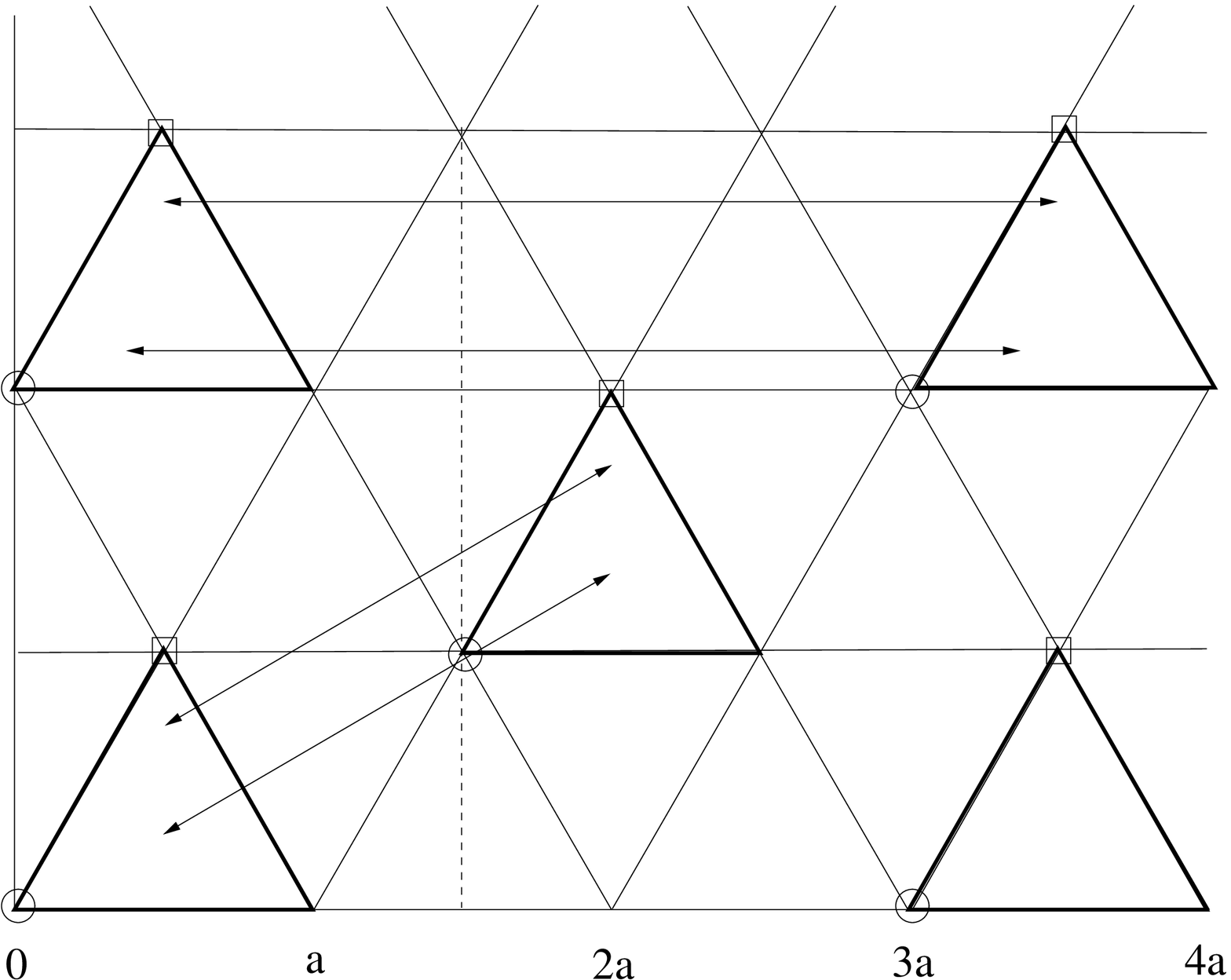,width=14cm,angle=0}
\caption{ }
\end{figure}

\newpage

\begin{figure}
\epsfig{file=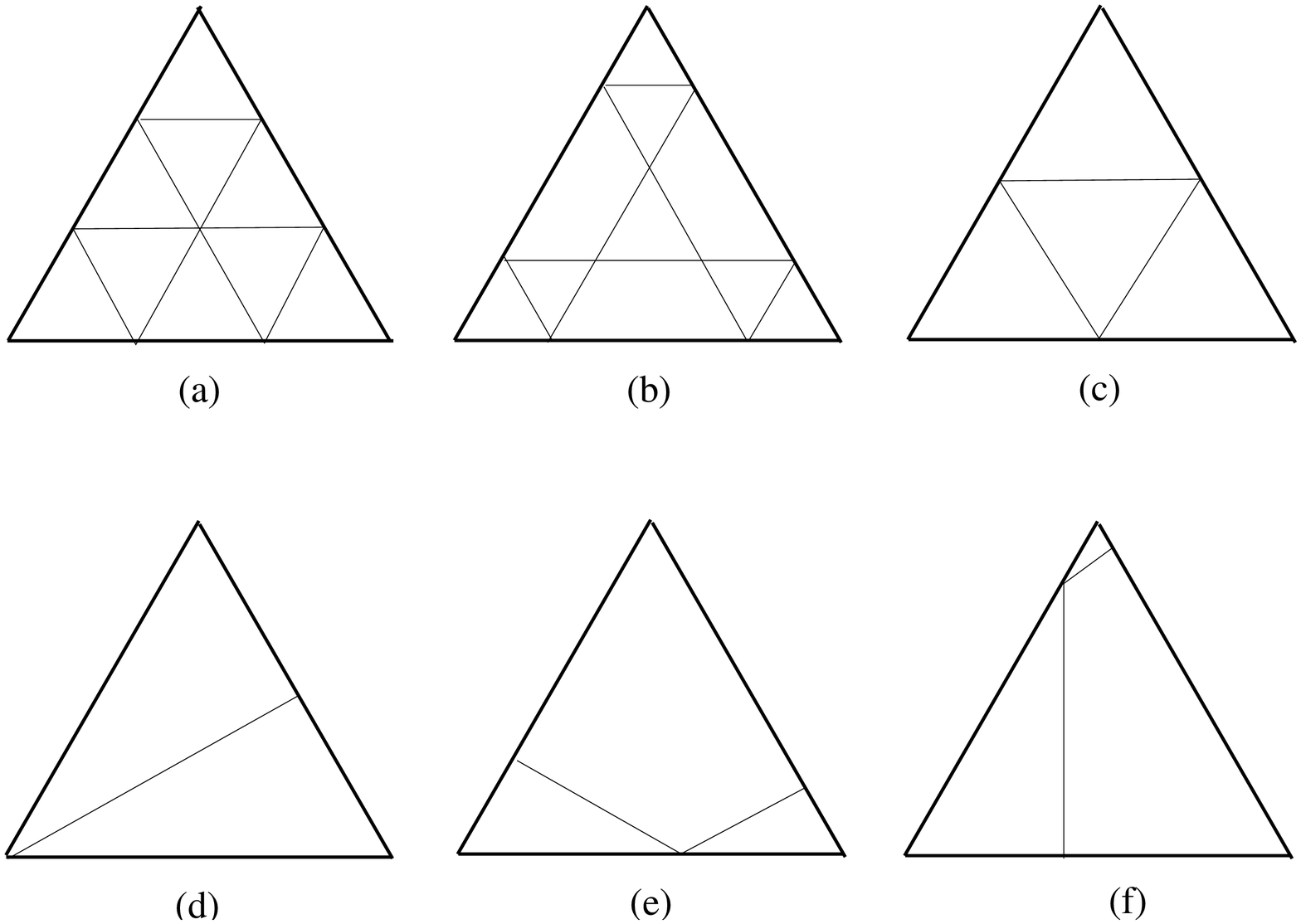,width=14cm,angle=0}
\caption{ }
\end{figure}

\newpage

\begin{figure}
\epsfig{file=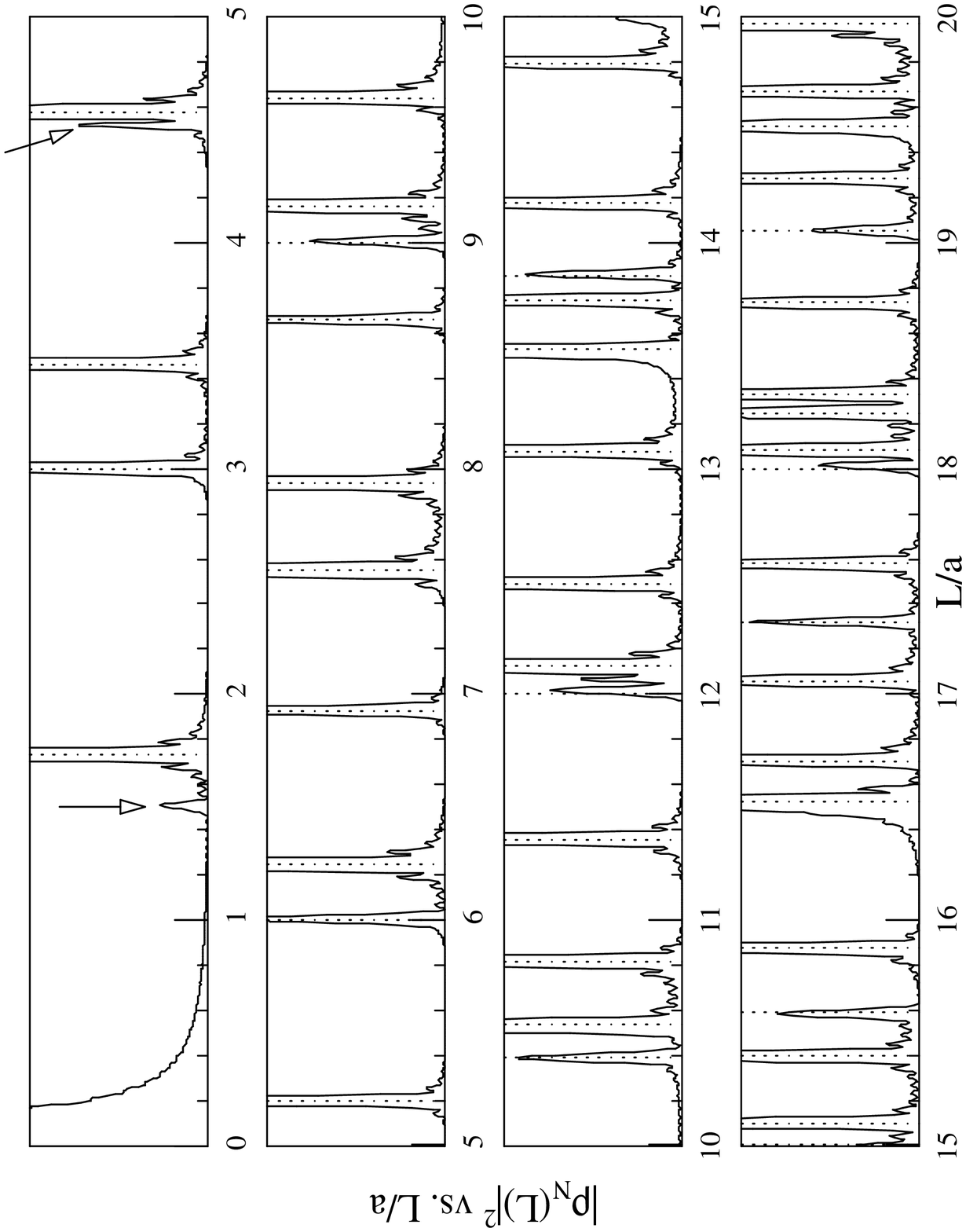,width=14cm,angle=0}
\caption{ }
\end{figure}

\newpage

\begin{figure}
\epsfig{file=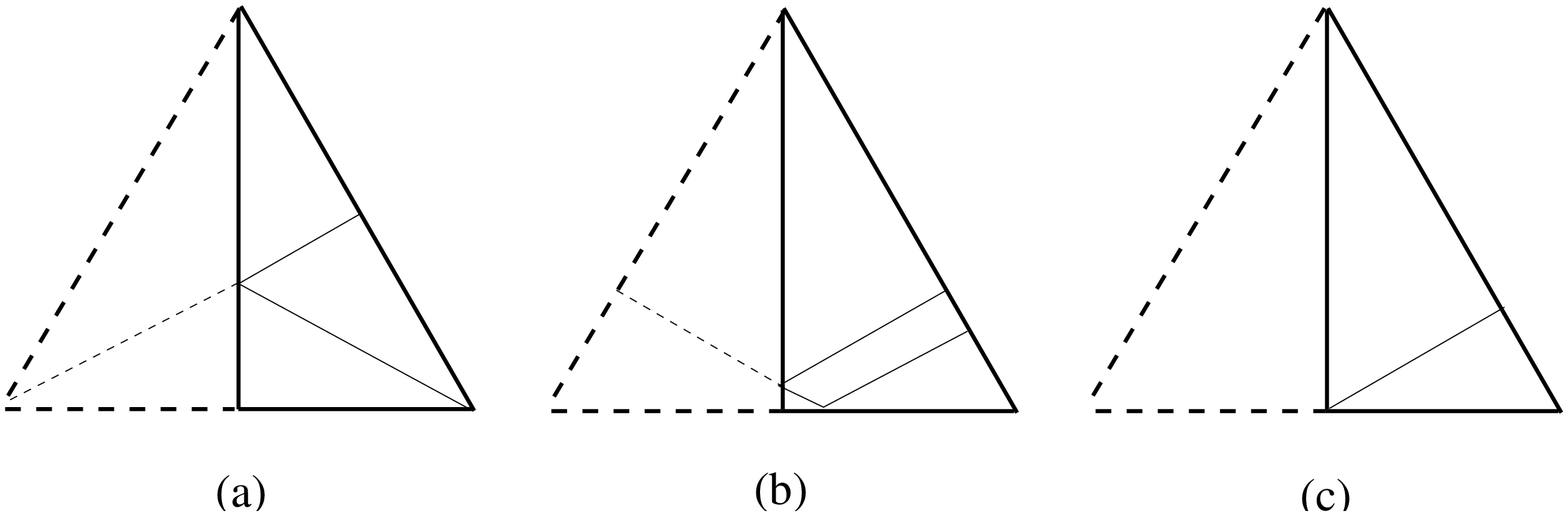,width=14cm,angle=0}
\caption{ }
\end{figure}

\newpage

\begin{figure}
\epsfig{file=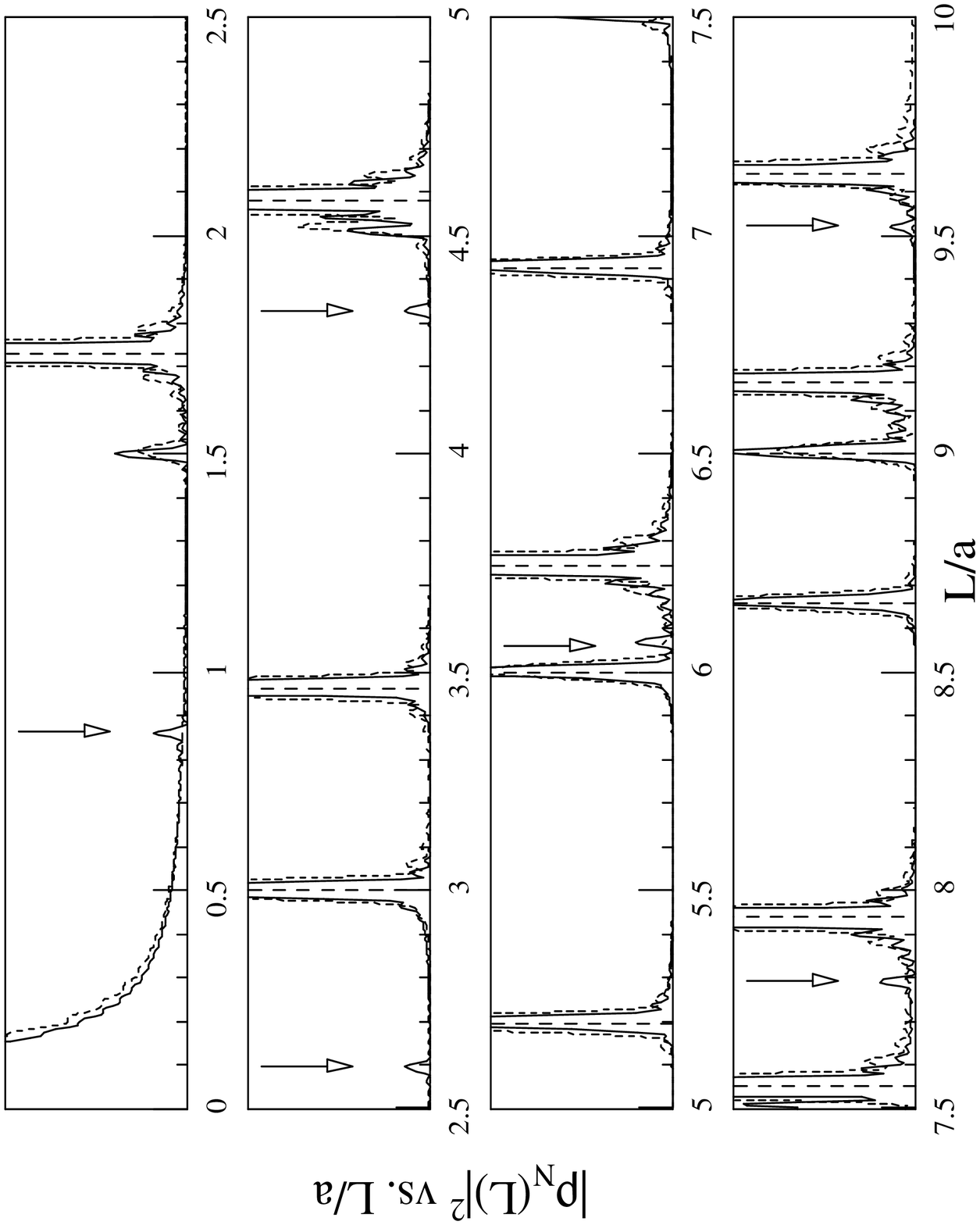,width=14cm,angle=0}
\caption{ }
\end{figure}

\newpage

\begin{figure}
\epsfig{file=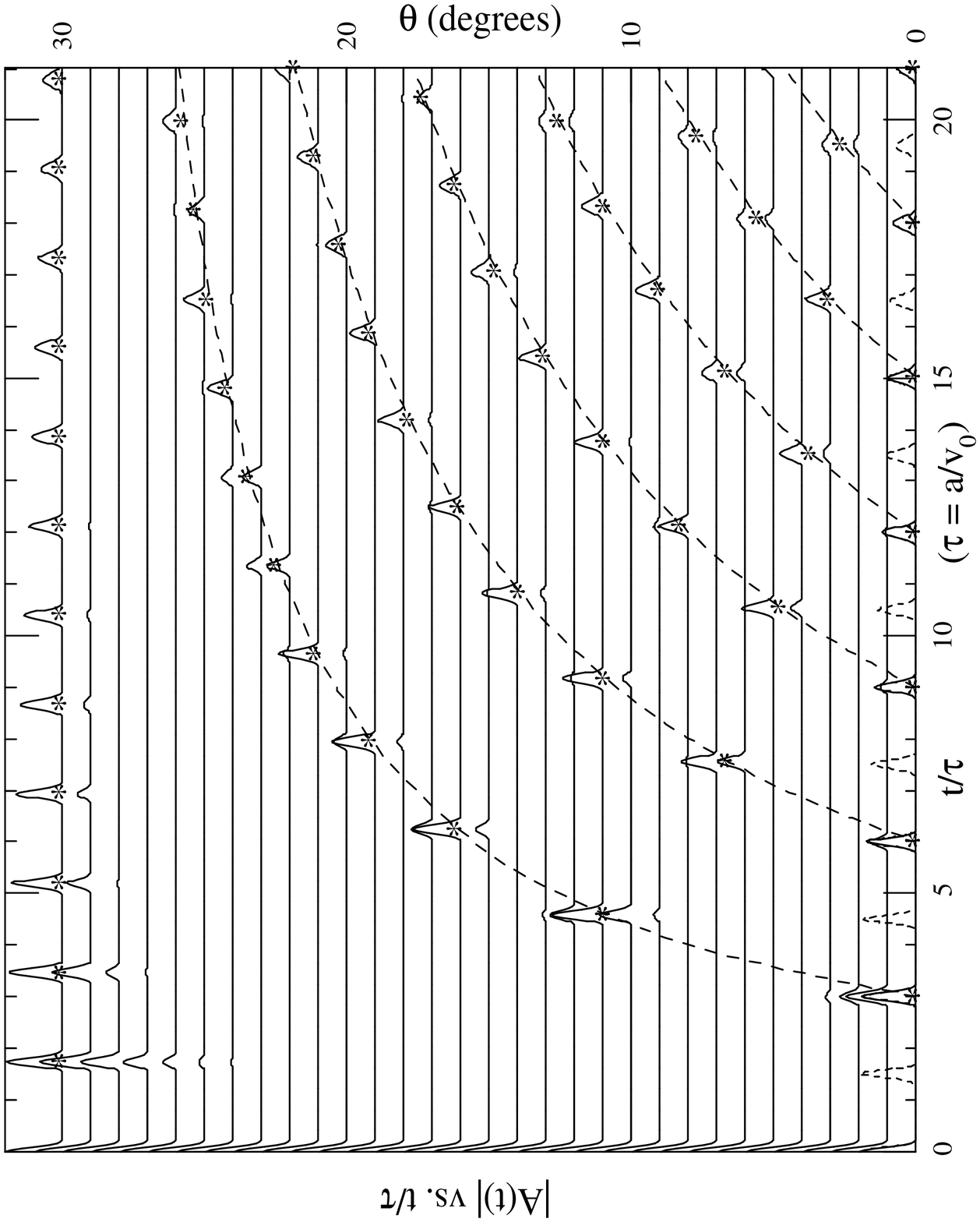,width=14cm,angle=0}
\caption{ }
\end{figure}

\newpage

\begin{figure}
\epsfig{file=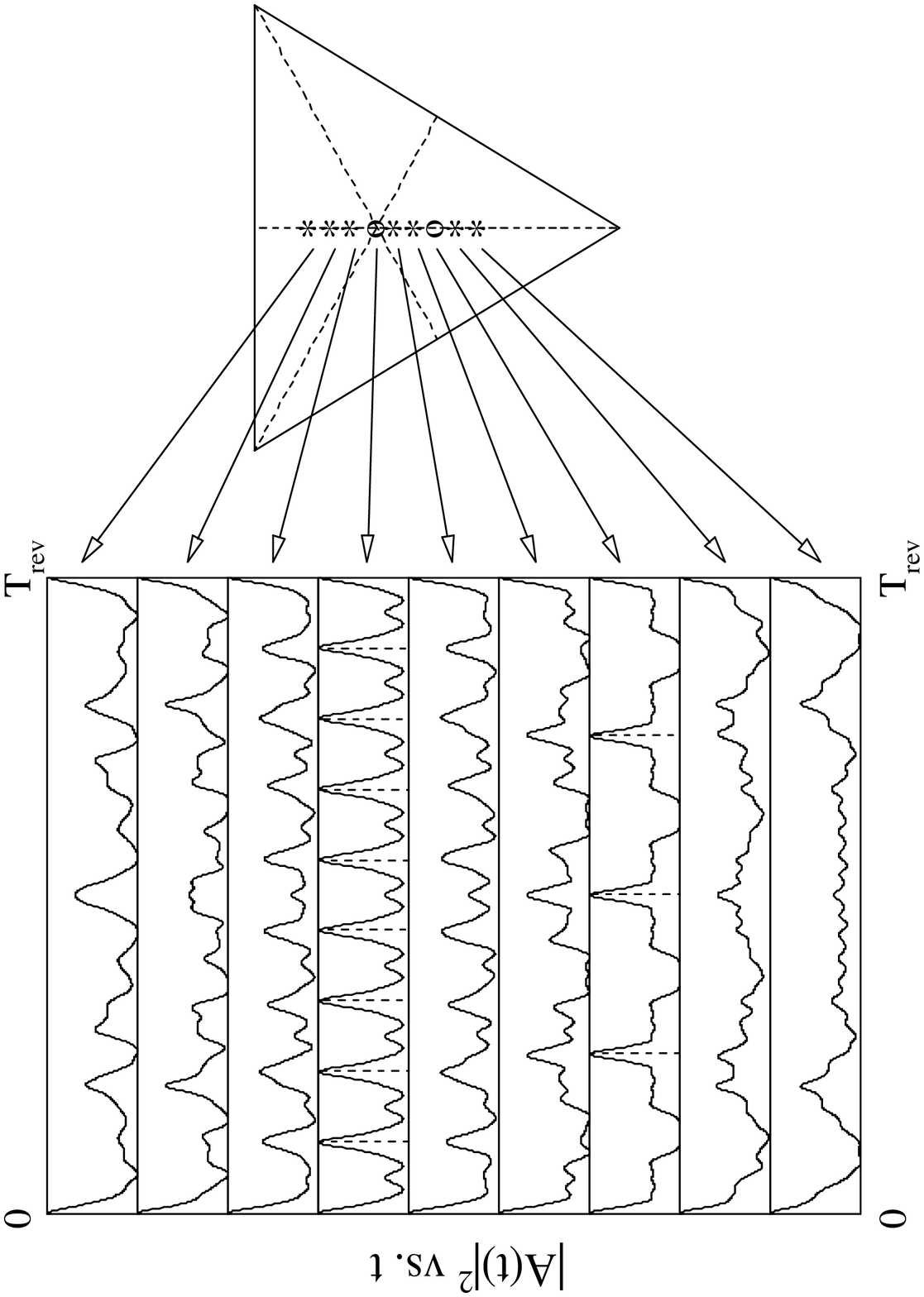,width=14cm,angle=0}
\caption{ }
\end{figure}

\end{document}